\documentclass[graybox]{svmult}

\usepackage{amsmath}
\usepackage{helvet}         % selects Helvetica as sans-serif font
\usepackage{courier}        % selects Courier as typewriter font
\usepackage{type1cm}        % activate if the above 3 fonts are
\usepackage{makeidx}         % allows index generation
\usepackage{graphicx}        % standard LaTeX graphics tool
\usepackage{multicol}        % used for the two-column index
\usepackage[bottom]{footmisc}% places footnotes at page bottom

\newcommand\be{\begin{equation}}
\newcommand\ee{\end{equation}}
\newcommand{\bea}{\begin{eqnarray}}
\newcommand{\eea}{\end{eqnarray}}

\newcommand{\nn}{\nonumber}
\newcommand{\pd}{\partial}

\begin{document}

\title*{On Non-slow Roll Inflationary Regimes}
\author{Lilia Anguelova, Peter Suranyi and L.C. Rohana Wijewardhana}
\institute{Lilia Anguelova \at Institute for Nuclear Research and Nuclear Energy, BAS, Sofia 1784, Bulgaria,\\ \email{anguelova@inrne.bas.bg}
\and Peter Suranyi \at Department of Physics, University of Cincinnati, OH, 45221, USA,\\ \email{peter.suranyi@gmail.com}
\and L.C. Rohana Wijewardhana \at Department of Physics, University of Cincinnati, OH, 45221, USA,\\ \email{rohana.wijewardhana@gmail.com}}
\maketitle

\abstract{We summarize our work on constant roll inflationary models. It was understood recently that constant roll inflation, in a regime beyond the slow roll approximation, {\it can} give models that are in agreement with the observational constraints. We describe a new class of constant roll inflationary models and investigate the behavior of scalar perturbations in them. We also comment on other non-slow roll regimes of inflation.}

\section{Introduction}
\label{sec:1}

It has long been a standard lore that, to agree with the observational constraints, an inflationary model has to be in the so called slow-roll regime. This is an approximation that allows an easy solution of the coupled equations of motion. The background metric is (near-)de Sitter and the spectrum of scalar perturbations turns out to be (nearly-)scale invariant, as required for consistency with the data from current cosmological observations. However, it is known since \cite{WK} that a scale invariant spectrum can also be obtained from a non-slow roll inflationary expansion. Although, the ultra-slow roll regime investigated in \cite{WK} is unstable (i.e. very short-lived) and thus cannot provide a full-fledged inflationary model by itself.  

Despite the stability issue, ultra-slow roll inflation has received considerable attention during the last several years in relation to the observed low-$l$ anomaly of the CMB \cite{CK}. It was also understood recently how to construct a class of ultra-slow roll composite inflation models in the context of the gauge/gravity duality \cite{LA,ASW,ASW_II}. Much more importantly, \cite{MSY} showed that a certain generalization of ultra-slow roll, called constant roll, can give a long-lasting/stable expansion in addition to producing a scale invariant spectrum of scalar perturbations. Therefore, constant roll inflation is an observationally viable alternative to the standard slow roll one.

In view of the great, and continually growing, precision of present day cosmological observations, it is undoubtedly worth investigating in more depth the full set of viable inflationary regimes. In \cite{ASW2} we performed a systematic study of the constant roll condition and found a new class of solutions of this type. These solutions are stable under scalar perturbations and have a corner of their parameter space, in which one obtains a nearly scale invariant spectrum of scalar perturbations. Here we discuss their properties and comment on broader non-slow roll regimes.

\section{Constant roll inflation}
\label{sec:2}

Within the standard field theoretic description, inflation is obtained as a solution of the equations of motion following from the action
\be
S = \int d^4 x \sqrt{-g} \left[ \frac{R}{2} + \frac{1}{2} g^{\mu \nu} \pd_{\mu} \phi \pd_{\nu} \phi - V(\phi) \right] \,\,,
\ee
upon using the metric ansatz
\be
ds_4^2 = -dt^2 + a^2 (t)\,d\vec{x}^2
\ee 
with $a(t)$ being the scale factor. 

The condition for inflationary solutions is $\ddot{a} (t) > 0$. In principle, such solutions may or may not satisfy the slow roll approximation, which can be defined in terms of the Hubble parameter \,$H(t) \equiv \frac{\dot{a}(t)}{a(t)}$ \,as \cite{OO,OO_II}:
\be \label{eta_phi}
\varepsilon \equiv - \frac{\dot{H}}{H^2} \,<\!\!< 1 \qquad {\rm and} \qquad \eta \equiv - \frac{\ddot{H}}{2 H \dot{H}} \,<\!\!< 1 \qquad .
\ee
The standard lore for decades has been that conditions (\ref{eta_phi}) are necessary in order to obtain a long-lasting (i.e. stable) inflationary expansion, which produces a scale invariant spectrum of scalar perturbations. In other words, it is usually assumed that (\ref{eta_phi}) is needed for consistency with the observational data.

However, it is also well-known that the ultra-slow roll regime of \cite{WK,TW}, defined by the conditions
\be
\varepsilon <\!\!< 1 \qquad {\rm and} \qquad \eta = 3 \quad \,\,\,\,\, ,
\ee
similarly gives a scale invariant spectrum, i.e. with $n_s=1$. This regime, though, is unstable and can last only a few e-folds. Recently \cite{MSY} showed that a generalization of the $\eta$-condition, given by
\be \label{CRC}
\eta = const \equiv c
\ee
and called constant roll regime, can lead to a long-lasting inflationary expansion, while preserving the $n_s = 1$ result, for some values of $c\neq 3$. The considerations of \cite{MSY} were based on the definition of the $\eta$-parameter as \,$\eta = -\frac{\ddot{\phi}}{H \dot{\phi}}$\,, \,which is equivalent to the one in (\ref{eta_phi}) upon using the field equations. A more straightforward and systematic analysis can be performed by studying instead the condition
\be \label{Heq}
- \frac{\ddot{H}}{2 H \dot{H}} = c \,\,\, ,
\ee
following from the $\eta$-definition in (\ref{eta_phi}) together with (\ref{CRC}).

Investigating equation (\ref{Heq}), the work \cite{ASW2} reproduced the solutions of \cite{MSY} and, in addition, found a new class of constant roll solutions. The Hubble parameter, scale factor and inflaton of the new solutions have the following form:
\bea \label{NSol}
H (t) \,&=& \,\frac{N}{c} \cot (Nt) \,\,\,\, , \nn \\
a (t) \,&=& \,C_a \sin^{1/c} (N t) \,\,\,\, , \nn \\
\phi (t) \,&=& \,\pm \sqrt{\frac{2}{c}} \ln \!\left[ \cot \!\left( \frac{Nt}{2} \right) \!\right] + C_{\phi} \,\,\,\, ,
\eea
where $N$, $C_a$ and $C_{\phi}$ are integration constants. Also, the parameter $c$ has to satisfy
\be \label{cpos}
c > 0 \,\,\, ,
\ee
to ensure $\dot{H} < 0$ (and thus a real inflaton $\phi$), while the combination $Nt$ has to be in the finite interval
\be \label{Ntint}
N t \in \left[ 0, \frac{\pi}{2} \right] \,\,\, ,
\ee
to have \,$H(t) > 0$ \,during the entire inflationary period. Clearly, by taking 
\be
N <\!\!< 1 \,\,\, ,
\ee
one can have as large a $t$-interval as desired.

Finally, the scalar potential is given by
\be \label{Vpot}
V(\phi) = \frac{N^2}{2 c^2} \left[ (3-c) \cosh \!\left( \sqrt{2c} \,(\phi + \phi_0) \right) - (3 + c) \right] \,\,\, ,
\ee
where the constant $\phi_0 \equiv - C_{\phi}$. We will see shortly that $V (\phi)$ is positive-definite within the entire inflationary parameter space of these constant roll models.

\section{Parameter space of the new solutions}
\label{sec:3}

The class of solutions (\ref{NSol}), with parameter space as in (\ref{cpos})-(\ref{Ntint}), was obtained only by studying the defining equation for constant roll, namely equation (\ref{Heq}), and imposing the requirements for a positive Hubble parameter and a real inflaton. However, we still need to consider the condition for inflation $\ddot{a} (t) > 0$\,. Now we will discuss the additional constraints on the parameter space of the new solutions that follow from this condition.

First, however, let us make an important observation. Note that the $Nt$-interval in equation (\ref{Ntint}) can be shortened by a rescaling of the integration constant $N$ \cite{ASW2}. Indeed, introducing the constant \,$\hat{N} = \frac{2}{\pi} \theta_* N$ \,with some fixed $\theta_* \!< \frac{\pi}{2}$\,\,, we can see that \,$Nt \in [0, \frac{\pi}{2}]$ \,becomes \,$\hat{N} t \in [0, \theta_*]$\,\,. So the freedom to redefine the integration constant $N$ implies that we are free to restrict the $Nt$-interval to a convenient subinterval. Clearly, this does not affect the above statement that the $t$-interval can be as large as desired, since the rescaled integration constant $\hat{N}$ is, obviously, just as arbitrary as $N$. However, it will be useful, at some point later on, to restrict the $Nt$ interval to $[0, \frac{\pi}{4}]$\,.

Now let us turn to investigating the condition $\ddot{a} > 0$\,. From (\ref{NSol}), we find:
\be \label{accel}
\ddot{a} (t) = \frac{N^2}{c^2} \frac{a (t)}{ \sin^2 ( N t ) } \left[ \,\cos^2 \!\left( N t \right) - c \,\right] \,\,\, .
\ee
Therefore, to ensure $\ddot{a} > 0$\,, one needs to satisfy the inequality 
\be \label{CondPosAc}
\cos^2 \!\left( N t \right) > c \quad. 
\ee
To be able to do that, we must have $c < 1$\,. Together with (\ref{cpos}), this implies that:
\be \label{c_par_space}
0 < c < 1 \quad .
\ee
Then we can solve (\ref{CondPosAc}), finding: 
\be \label{Ntmax_c}
Nt \in \left[\,0, \,{\rm arccos (\sqrt{c})}\,\right) \,\,\, .
\ee
Note that (\ref{Ntmax_c}) guarantees the positive-definiteness of the inflaton potential (\ref{Vpot}); see \cite{ASW2}.

Finally, let us discuss what are the conditions for the acceleration in the new class of models to be increasing or decreasing. Computing the time-derivative of (\ref{accel}), we have:
\be
%\frac{d\ddot{a}}{dt}
\dddot{a} = \frac{N^2}{c^2} \frac{a H}{\sin^2 (N t)} \left[ \cos^2 (N t) - 3c + 2c^2 \right] \,\,\, .
\ee
Hence, the condition $\dddot{a} > 0$
%$\frac{d\ddot{a}}{dt} >0$ 
is equivalent to
\be \label{Condaccelpos}
\cos^2 (N t) > 3c - 2c^2 \,\,\, .
\ee
Note that, when $\frac{1}{2} < c < 1$\,, one always has $3c - 2c^2 > 1$ \,. So, in that case, $\ddot{a} (t)$ is always decreasing with time. On the other hand, when $c < \frac{1}{2}$\,, one can solve the condition for increasing acceleration (\ref{Condaccelpos}), obtaining:
\be
Nt < {\rm arccos} \left( \!\sqrt{3 c - 2 c^2} \right) \,\,\, .
\ee
In conclusion, to have any period of increasing acceleration (like in the familiar de Sitter case), one has to have \,$c < \frac{1}{2}$\,\,.

\section{Stability under scalar perturbations}
\label{sec:4}

Let us now discuss the scalar perturbations in the new class of models (\ref{NSol}) with parameter space (\ref{Ntint}) and (\ref{c_par_space}). We will denote the perturbations of the inflaton and the spatial part of the metric as $\delta \phi$ and $\delta g_{ij}$ respectively, where $i,j=1,2,3$. It is convenient to work in comoving gauge, where $\delta \phi = 0$ and $\delta g_{ij} = a^2 \left[ ( 1 - 2 \zeta) \delta_{ij} + h_{ij} \right]$ with $h_{ij}$ being the tensor perturbations; see \cite{DB} for instance. As is well-known, the perturbation $\zeta$ inside $\delta g_{ij}$ is the only independent scalar degree of freedom.

Upon Fourier transforming $\zeta (t,\vec{x}) = \int \!\frac{d^3 k}{(2\pi)^3} \,\zeta_k (t) \,e^{i \vec{k}.\vec{x}}$\,, one can introduce the mode function $v_k \equiv \sqrt{2} \,z \zeta_k$ with $z^2 \equiv - a^2 \!\frac{\dot{H}}{H^2}$\,. In terms of $v_k$\,, the evolution equation for the perturbations is the Mukhanov-Sasaki equation \cite{VM,MSa}: 
\be \label{MukhSasEq}
v_k'' + \left( k^2 - \frac{z''}{z} \right) v_k = 0 \,\,\, ,
\ee
where $k \equiv |\vec{k}|$ and $'\!\equiv \!\pd_{\tau}$ with $\tau$ being conformal time defined as usual via \,$dt^2 = a^2 d\tau^2$\,. Note also that the $z''/z$ term in (\ref{MukhSasEq}) can be rewritten {\it exactly} (as opposed to in the slow-roll approximation) as \cite{LL,MSY}:
\be \label{MassTerm}
\frac{\tilde{z}''}{\tilde{z}} = a^2 H^2 \left( 2 - \epsilon_1 + \frac{3}{2} \epsilon_2 + \frac{1}{4} \epsilon_2^2 - \frac{1}{2} \epsilon_1 \epsilon_2 + \frac{1}{2} \epsilon_2 \epsilon_3 \right) \,\, ,
\ee
where $\epsilon_i$ are the following series of slow roll parameters:
\be \label{srPar}
\epsilon_1 \equiv - \frac{\dot{H}}{H^2} \qquad {\rm and} \qquad \epsilon_{i+1} \equiv \frac{\dot{\epsilon}_i}{H \epsilon_i} \quad .
\ee

To investigate the issue of stability of the new models under scalar perturbations, we will consider the super-Hubble limit of the evolution equation (\ref{MukhSasEq}), where $k^2 <\!\!< \tilde{z}''\!/\tilde{z}$\,. Clearly, in that case, (\ref{MukhSasEq}) simplifies to:
\be \label{MukhSas}
v_k'' - \frac{\tilde{z}''}{\tilde{z}} \,v_k = 0 \,\,\, .
\ee
It was already observed in \cite{MSY} that the general solution of (\ref{MukhSas}) gives the following form for $\zeta_k = \frac{\sqrt{2}}{2} \frac{v_k}{\tilde{z}}$\,:
\be \label{zeta}
\zeta_k = A_k + B_k \!\int \!\!\frac{dt}{a^3 \epsilon_1} \,\,\, ,
\ee
where $A_k,B_k=const$ and $\tau = \tau (t)$ is any function. Using (\ref{srPar}) and absorbing a minus sign in the arbitrary integration constant $B_k$, we can conveniently rewrite (\ref{zeta}) as:
\be \label{zetaHdot}
\zeta_k = A_k + B_k \!\int \!\!\frac{\,H^2}{a^3 \dot{H}} \, dt \quad .
\ee
The goal now will be to investigate the behavior of this integral at late times. If it turns out that $\zeta_k$ decreases (or stays constant), then the corresponding model would be stable. On the other hand, if it were to increase with time, this would indicate instability. 

Substituting the Hubble parameter $H$ and scale factor $a$ from (\ref{NSol}), we find:
\bea \label{Icase3}
\int \!\frac{H^2}{a^3 \dot{H}} \,dt 
&=& \,\frac{1}{3 \,c \,N (C_a)^3} \,\cos^3 (Nt) \,\, {}_2F_1 \!\left( \frac{3}{2} \,, \frac{c+3}{2c}\,, \frac{5}{2} \,; \,\cos^2 (Nt) \!\right) \,\, .
\eea
Note that the parameter $c$ here and the parameter $\alpha$ in \cite{MSY} are related to each other via $c = 3 + \alpha$. Using this, one can immediately verify that the indices of the hypergeometric function in (\ref{Icase3}) coincide precisely with those in eq. (47) of \cite{MSY}. In fact, if we denote $x \equiv \cos^2 (Nt)$\,, we have in (\ref{Icase3}) exactly the same function (up to an overall numerical constant) as in eq. (47) of \cite{MSY}, namely 
\be \label{fx}
f (x) \equiv x^{\frac{3}{2}}\, {}_2F_1 \!\left( \,\frac{3}{2} \,, \frac{c+3}{2c}\,, \frac{5}{2} \,; \,x \right) \,\,\, .
\ee
However, there is a crucial difference due to the fact that reference \cite{MSY} needed to investigate that function in the limit $x \rightarrow \infty $, whereas for us $x \in [c,1]$ because of (\ref{Ntmax_c}).

\begin{figure}[t]
\begin{center}
\hspace*{-0.2cm}\includegraphics[scale=0.29]{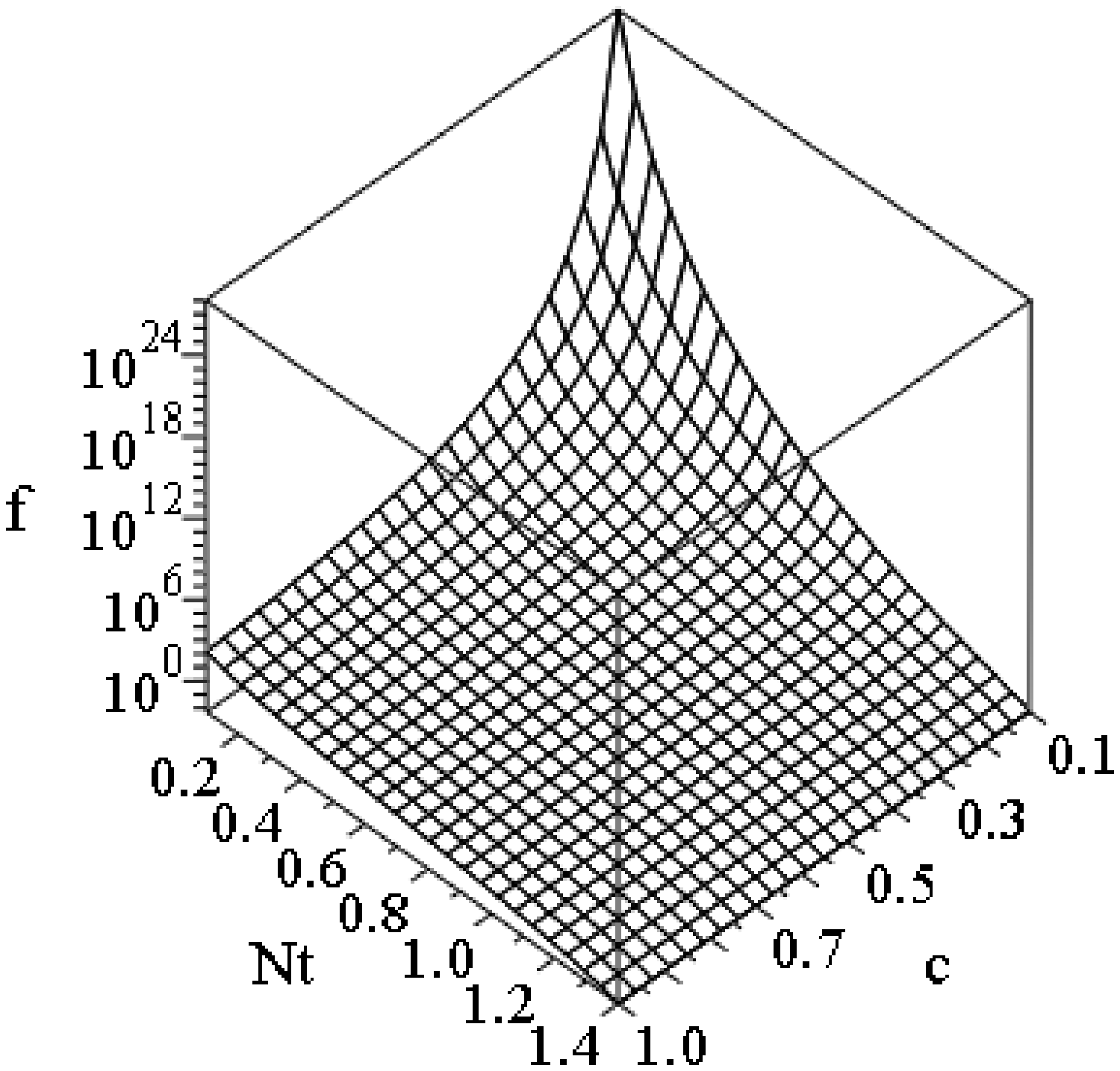}
\hspace*{-0.4cm}\includegraphics[scale=0.33]{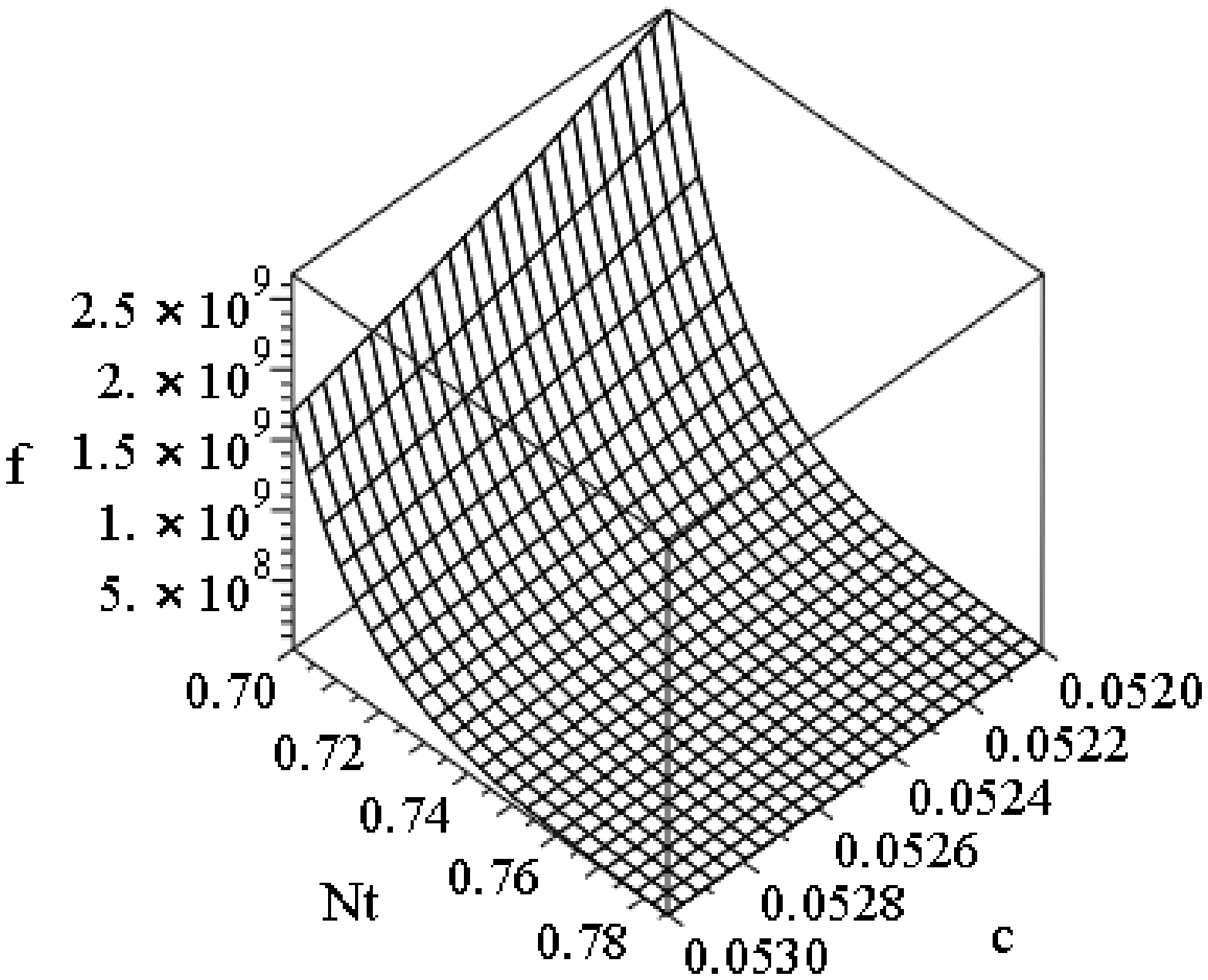}
\end{center}
\vspace{-0.2cm}
\caption{{\small Plot of $f(x)$ in equation (\ref{fx}), with $x=\cos^2 (Nt)$\,, as a function of both $Nt$ and $c$. On the left side, $f(Nt,c)$ is plotted for $Nt\in (0,\frac{\pi}{2})$ and $c\in (0,1)$\,. On the right side, we have plotted a representative slice for the intervals $Nt \in [\,0.7\,,\,\frac{\pi}{4}\,]$ and $c \in [\,0.053\,,\,0.054\,]$\,, which will be useful in the next Section.}}
\label{3dPlots}
\vspace{0.3cm}
\end{figure}

In \cite{ASW2} the function (\ref{fx}) was considered in the full parameter ranges, given by   $0<Nt<\frac{\pi}{2}$ and $0<c<1$\,, and it was shown that it is always decreasing with time regardless of the values of the constants $c$ and $N$. This behavior is illustrated on Figure \ref{3dPlots}. 

\section{Scalar spectral index}
\label{sec:5}

In order to determine the scalar spectral index $n_s$\,, we need to investigate the Mukhanov-Sasaki equation (\ref{MukhSasEq}) in a regime when the terms with $k^2$ and $\frac{z''}{z}$ are comparable. We will impose the usual initial condition:
\be \label{initial_cond}
v_k (\tau) = \frac{e^{-i k \tau}}{\sqrt{2 k}} \qquad {\rm for} \qquad \tau \rightarrow -\infty \quad .
\ee

To make further progress, we need the explicit relation $\tau = \tau (t)$. Using $a(t)$ in (\ref{NSol}), one finds:
\be
\tau = \int \!\frac{dt}{a} \,= \,- \,\frac{1}{C_a N} \,\,\cos (Nt) \,\,{}_2 F_1 \!\left( \frac{1}{2}\,, \,\frac{c+1}{2c}\,, \,\frac{3}{2}\,; \,\cos^2 (Nt) \!\right) \,+ \,\,const \,\,\, .
\ee
The integration constant here can easily be chosen such that the range of $\tau$ is \cite{ASW2}:
\be \label{tau_range}
\tau \in (-\infty \,, \,0\,]
\ee
for $t$ varying in the entire interval
\be \label{posac_t_int}
t \in \left[ 0,\,\frac{1}{N} \arccos (\sqrt{c}) \!\right) \,\, ,
\ee
according to the inflationary condition (\ref{Ntmax_c}). 

As discussed in Section \ref{sec:3} though, we can restrict to any subinterval of (\ref{posac_t_int}) as part of the freedom to redefine the integration constant $N$. It will turn out below to be particularly useful to consider the subinterval
\be \label{t_int_Pi_o_4}
t \in \left[ 0, \frac{\pi}{4N}\right] \,\,\, 
\ee
when $c < \frac{1}{2}$\,. In this case, the integration constant guaranteeing (\ref{tau_range}) is such that:
\bea \label{tau_function}
\tau \,&=& \,- \,\frac{1}{C_a N} \left[ \cos (Nt) \,\,{}_2 F_1\!\left( \frac{1}{2}\,, \,\frac{c+1}{2c}\,, \,\frac{3}{2}\,; \,\cos^2 (Nt) \!\right) \right. \nn \\
&-& \left. \frac{\sqrt{2}}{2} \,\,{}_2 F_1\!\left( \frac{1}{2}\,, \,\frac{c+1}{2c}\,, \,\frac{3}{2}\,; \,\frac{1}{2} \right) \right] \,\,\, .
\eea

Solving equation (\ref{MukhSasEq}) in full generality is rather complicated because the potential term $z''/z$ depends on the background. In principle, one needs to use numerical methods \cite{DB}. However, one can find an analytical estimate, compatible with the observational constraint $n_s \approx 1$\,, in the approximation
\be \label{cless1}
c <\!\!< 1 \,\,\, .
\ee
In this limit, we are free to choose the interval (\ref{t_int_Pi_o_4}) as our inflationary period. And, furthermore, during that entire period the slow roll parameters $\epsilon_i$ in (\ref{srPar}) are almost constant \cite{ASW2}. More concretely, we have:
\be \label{epsilon_i_approx}
\epsilon_1 \approx 2 c \quad \,\,, \quad \,\, \epsilon_2 \approx 2 c \quad \,\,, \quad \,\, \epsilon_3 \approx 4 c \quad .
\ee
Hence, the $\epsilon_i$-expression in (\ref{MassTerm}) acquires the form:
\be \label{Expr_epsilons}
\left( 2 - \epsilon_1 + \frac{3}{2} \epsilon_2 + \frac{1}{4} \epsilon_2^2 - \frac{1}{2} \epsilon_1 \epsilon_2 + \frac{1}{2} \epsilon_2 \epsilon_3 \right) \approx \,2 + c + 3 c^2 = \nu^2 - \frac{1}{4} \,\,\,\, ,
\ee
where for convenience we have introduced the notation
\be
\nu^2 \equiv \frac{9}{4} + c + 3 c^2 \,\,\,\, .
\ee
In addition, in the approximation (\ref{cless1}), one can verify from (\ref{NSol}) and (\ref{tau_function}) that \cite{ASW2}:
\be \label{ah_inT}
a H \approx - \frac{1}{\tau} \,\,\,\, ,
\ee
similarly to inflation in pure de Sitter space. Now, making use of (\ref{Expr_epsilons}) and (\ref{ah_inT}) inside (\ref{MukhSasEq}), one can easily obtain the spectral index $n_s$ by following the standard computation \cite{ASW2,MSY}. The result is:
\be \label{n_s_expr}
n_s = 4 - 2 \sqrt{\frac{9}{4} + c + 3 c^2} \,\,\,\, .
\ee

To find the values of $c$ that lead to agreement with the observational constraint $n_s \approx 0.96$\,, we need to solve the quadratic equation that follows from imposing it on equation (\ref{n_s_expr}). It turns out that only one of the two roots lies within the parameter space of our class of models, namely within (\ref{c_par_space}). That solution is:
\be
c \approx 0.0522 \,\,\,\, .
\ee
As explained in \cite{ASW2}, this result is consistent with the approximations made in deriving it. More precisely, for this value of the parameter $c$, the approximations (\ref{epsilon_i_approx}) and (\ref{ah_inT}) hold to a very good degree of accuracy. Hence, we have found a corner of the parameter space of the new constant roll models, in which they are compatible with the present day observational data.

\section{Other non-slow roll regimes}
\label{sec:6}

The constant roll regime studied here can be viewed as a generalization of ultra-slow roll inflation, that was first considered in \cite{TW}. Other non-slow roll inflationary regimes have also been investigated during the last couple of decades. See, for example \cite{AL,CK}, for different cases of `fast roll' inflation, depending on which (and how many) of the slow roll parameters in (\ref{srPar}) are actually large during the inflationary period. Usually, such stages of expansion are expected to be rather short-lived. So they are viewed as useful only for setting certain initial conditions for a subsequent stage of regular slow-roll inflation. A transient non-slow roll stage preceding slow roll is, in fact, considered to be important in explaining the observed low multipole moment anomaly in the CMB \cite{CK}. 

However, in view of the recent realization \cite{MSY,ASW2}, that a constant roll inflationary expansion can last long enough to produce a full-fledged inflationary model (compatible with $n_s \approx 1$ in a part of its parameter space), it makes sense to ask whether it is possible to find other stable non-slow roll regimes. In particular, it would be interesting to investigate whether there is a suitable generalization of fast roll inflation, conceptually similar to how constant roll generalizes ultra-slow roll. We hope to come back to this question in the future.

Finally, in view of the fact that, at present, the Universe has a (small) positive cosmological constant, it is worth exploring models that can have more than one inflationary stage. This would enable the development of a unified description, that can account for {\it both} inflation in the Early Universe {\it and} accelerated expansion in the present day. Important progress in that direction was achieved in \cite{OOS}. The early inflationary period in their considerations was with constant rate of roll. It would be interesting to explore how our new constant-roll solutions fit in this framework and, in particular, whether they can lead to some specific observational features in this context.

\begin{acknowledgement}
L.A. has received partial support from the Bulgarian NSF grant DN 08/3 and the bilateral grant STC/Bulgaria-France 01/6.
\end{acknowledgement}

\end{document}